\documentclass{aa}  

\usepackage{hyperref}

\usepackage{amsmath}
\usepackage{graphicx}
\usepackage{graphics}
\usepackage{subcaption}
\usepackage{natbib}{}
\usepackage{xcolor}
\usepackage{url}
\usepackage{hyperref}

\bibpunct{(}{)}{;}{a}{}{,}

\def\teff{\ifmmode T_{\rm eff} \else $T_{\mathrm{eff}}$\fi}

\def\ltsima{$\buildrel<\over\sim$}
\def\lsim{\lower.5ex\hbox{\ltsima}}

\newcommand{\ha}{\ifmmode {\rm H}\alpha \else H$\alpha$\fi}
\newcommand{\hb}{\ifmmode {\rm H}\beta \else H$\beta$\fi}
\newcommand{\hg}{\ifmmode {\rm H}\gamma \else H$\gamma$\fi}
\newcommand{\lya}{\ifmmode {\rm Ly}\alpha \else Ly$\alpha$\fi}

\newcommand{\ebv}{\ifmmode E_{\rm B-V} \else $E_{\rm B-V}$\fi}
\newcommand{\av}{\ifmmode A_{\rm V} \else $A_{\rm V}$\fi}

\def\cmc{cm$^{-3}$}

\def\msun{\ifmmode M_{\odot} \else M$_{\odot}$\fi}
\def\msunyr{\ifmmode M_{\odot} {\rm yr}^{-1} \else M$_{\odot}$ yr$^{-1}$\fi}
\def\zsun{\ifmmode Z_{\odot} \else Z$_{\odot}$\fi}

\def\lsun{\ifmmode L_{\odot} \else L$_{\odot}$\fi}

\def\mup{\ifmmode M_{\rm up} \else M$_{\rm up}$\fi}
\def\mlow{\ifmmode M_{\rm low} \else M$_{\rm low}$\fi}
\def\aap{A\&A}

\def\apjl{ApJL}
\def\apjs{ApJS}

\newcommand{\oh}{\ifmmode 12 + \log({\rm O/H}) \else$12 + \log({\rm
O/H})$\fi}
\def\Nii{[N\small II]\normalsize $\lambda\lambda$6584}

\def\Oii{[O~{\sc ii}] $\lambda$3727}

\def\Oiiit{[O~{\sc iii}]$\lambda 4363$}

\def\flyf{\ifmmode f_{\rm Lyf} \else $f_{\rm Lyf}$\fi}
\def\pz{\ifmmode P(z) \else $P(z)$\fi}
\def\ki2{\ifmmode \chi^2 \else $\chi^2$\fi}
\def\zphot{\ifmmode z_{\rm phot} \else $z_{\rm phot}$\fi}

\newcommand{\xphot}{\ifmmode x_\gamma \else $v_\gamma$\fi}
\newcommand{\xobs}{\ifmmode x_{\rm obs} \else $x_{\rm obs}$\fi}
\newcommand{\xcmf}{\ifmmode x_{\rm CMF} \else $x_{\rm CMF}$\fi}
\newcommand{\vexp}{\ifmmode V_{\rm exp} \else $V_{\rm exp}$\fi}
\newcommand{\vmax}{\ifmmode V_{\rm max} \else $V_{\rm max}$\fi}
\newcommand{\nh}{\ifmmode N_{\rm HI} \else $N_{\rm HI}$\fi}
\newcommand{\dv}{\ifmmode \Delta v({\rm em-abs}) \else $\Delta v({\rm em}-{\rm abs})$\fi}

\def\fesc{\ifmmode f_{\rm esc} \else $f_{\rm esc}$\fi}
\def\fescrel{\ifmmode f_{\rm esc,rel} \else $f_{\rm esc,rel}$\fi}

\def\frellya{\ifmmode f^{\rm rel}_{\rm{Ly}\alpha} \else $f^{\rm rel}_{\rm{Ly}\alpha}$\fi}

\newcommand{\mstar}{\ifmmode M_\star \else $M_\star$\fi}
\newcommand{\muv}{\ifmmode M_{\rm UV} \else $M_{\rm UV}$\fi}
\newcommand{\auv}{\ifmmode A_{\rm UV} \else $A_{\rm UV}$\fi}
\newcommand{\luv}{\ifmmode L_{\rm UV} \else $L_{\rm UV}$\fi}
\newcommand{\lir}{\ifmmode L_{\rm IR} \else $L_{\rm IR}$\fi}
\newcommand{\lbol}{\ifmmode L_{\rm bol} \else $L_{\rm bol}$\fi}
\newcommand{\liruv}{\ifmmode L_{\rm IR+UV} \else $L_{\rm IR+UV}$\fi}
\newcommand{\liroveruv}{\ifmmode L_{\rm IR}/L_{\rm UV} \else $L_{\rm IR}/L_{\rm UV}$\fi}
\newcommand{\nlyc}{\ifmmode N_{\rm Lyc} \else $N_{\rm Lyc} $\fi}
\newcommand{\rholyc}{\ifmmode \rho_{\rm Lyc} \else $\rho_{\rm Lyc} $\fi}
\newcommand{\chion}{\ifmmode \xi_{\rm ion} \else $\xi_{\rm ion}$\fi}
\newcommand{\chioncorr}{\ifmmode \xi_{\rm ion}^0 \else $\xi_{\rm ion}^0$\fi}

\newcommand{\Niiiuv}{N~{\sc iii}] $\lambda$1750}
\newcommand{\Nivuv}{N~{\sc iv}] $\lambda$1486}
\newcommand{\Nivuvab}{N~{\sc iv}] $\lambda$$\lambda$ 1483,1486}

\newcommand{\rhostar}{\ifmmode \rho_\star \else $\rho_\star$\fi}
\newcommand{\rhosfr}{\ifmmode \rho_{\rm SFR} \else $\rho_{\rm SFR}$\fi}
\newcommand{\rhogc}{\ifmmode \rho_{\rm GC} \else $\rho_{\rm GC}$\fi}
\newcommand{\zmin}{\ifmmode z_{\rm min} \else $z_{\rm min}$\fi}

\begin{document}

\title{N-emitters as possible signposts of GC formation}
\subtitle{}
\author{D. Schaerer\inst{1},
\thanks{E-mail: daniel.schaerer@unige.ch}, 
R. Marques-Chaves\inst{1},
H. Atek\inst{2},
N. Prantzos\inst{2},
C. Charbonnel\inst{1},
M. Talia\inst{3},
I. Morel\inst{1},  
M. Dessauges-Zavadsky\inst{1},
Y.I. Izotov\inst{4},
N. Guseva\inst{4}
}
  \institute{Observatoire de Gen\`eve, Universit\'e de Gen\`eve, Chemin Pegasi 51, 1290 Versoix, Switzerland
\and Institut d'Astrophysique de Paris, UMR 7095 CNRS, Sorbonne Universit\'e, 98bis, Bd Arago, 75014 Paris, France
\and University of Bologna, Department of Physics and Astronomy, Via Gobetti 93/2, 40129 Bologna, Italy
\and Bogolyubov Institute for Theoretical Physics,
National Academy of Sciences of Ukraine, 14-b Metrolohichna str., Kyiv,
03143, Ukraine
}

\authorrunning{Schaerer et al.}
\titlerunning{N-emitters as possible signposts of GC formation}

\date{Received date; accepted date}

\abstract{Based on the finding of unusual chemical abundance ratios of N-emitters, which resemble those of globular cluster (GC) stars, their compactness, high ISM densities and other properties, it has  been suggested that N-emitters could indicate the formation sites of globulars. 
A recent statistical study of the N-emitter population has quantified the frequency $f_N$ of these rare objects and their redshift evolution \citep{Morel2025Discovery-of-ne}. Using these results we here test if N-emitters trace the formation of GCs and use the observed cosmic star-formation rate density evolution to predict the cosmological evolution of the GC population with time, their age distribution, and the total present-day stellar mass density formed in globulars.
The predicted age distribution of GCs strongly resembles the typical asymmetric observed distributions in the Galaxy and ellipiticals, with a peak at $\sim 11.5-12$ Gyr and a longer tail extending to younger ages. 
We derive a total stellar mass density formed in N-emitters down to redshift zero of $(2-4) \times 10^5$ \msun\ Mpc$^{-3}$, which matches within a factor $\sim 2$ the observed fraction of stellar mass found in the GC population at $z=0$. 
These results provide additional indirect arguments supporting 
that N-emitters could represent signposts of a short phase of GC formation.}


\maketitle

\section{Introduction}
\label{s_intro}

Following up on pioneering works using strong gravitational lensing to identify high-redshift star cluster candidates \citep[e.g.][]{Vanzella2017Paving-the-way-,Vanzella2019Massive-star-cl}, deep high-resolution observations with the James Webb Space Telescope (JWST), are starting to provide new insights on the formation of the first star clusters, cluster complexes, and galaxies formed shortly after the Big Bang, including possibly the first {\em in situ} views of globular clusters (GCs) in formation \citep[e.g.][]{Vanzella2023JWST/NIRCam-Pro,Adamo2024Bound-star-clus,Claeyssens2025Tracing-star-fo}. 

Long thought to be the oldest and simple stellar systems, GCs are now well known to host multiple stellar populations with systematic star-to-star abundance variations, whose nature remains puzzling \citep[e.g.][]{Bastian2018Multiple-Stella}.

The recent discovery with JWST of a new class of rare objects with unusual nitrogen emission lines in the ultra-violet (UV), the so-called N-emitters, has opened a possibility to study GC formation at high-redshift.  Indeed, these objects show super-solar N/O abundance ratios at low metallicities 
\citep[e.g.][]{Bunker2023JADES-NIRSpec-S,Ji2026Connecting-JWST} and other properties resembling those of GCs, which has led several authors to suggest that N-emitters could be signposts of GCs in formation \citep{Charbonnel2023N-enhancement-i,Senchyna2024GN-z11-in-Conte,Marques-Chaves2024Extreme-N-emitt,Ji2026Connecting-JWST}.

To test this hypothesis and further explore the possible link between N-emitters and GCs, we here use the recent results from \cite{Morel2025Discovery-of-ne}, who provided the first systematic search and statistical study of the population of N-emitters using a large fraction of the available JWST spectra. Assuming that N-emitters trace the formation of GCs, we combine the observed fraction of N-emitters as a function of redshift, with recent measurements of the cosmic star-formation rate density evolution to predict the formation rate, age distribution, and total amount of stars formed in GCs, which we compare with observations. 

Our simple model and the derived predictions are presented and compared to observations in Sect.~\ref{s_gc}. We then discuss  implications and the main caveats (Sect.~\ref{s_discuss}), and summarise our results in Sect.~\ref{s_conclude}. 
We assume the following cosmology:  $\mathrm{\Omega_{m}=0.3}$, $\mathrm{\Omega_{\Lambda}=0.7}$, $\mathrm{H_{0}=70 \ km\ s^{-1} Mpc^{-1}}$, and a Chabrier IMF \citep{Chabrier2003Galactic-Stella}.

\section{From N-emitters to GC populations}
\label{s_gc}

\subsection{N-emitter statistics to predict the average cosmological GC population}
\label{s_stats} 
\cite{Morel2025Discovery-of-ne} have recently undertaken a statistical analysis of N-emitters using the low-resolution PRISM spectra obtained with the NIRSpec multi-object spectrograph onboard JWST. Searching for the presence of UV nitrogen emission lines of \Niiiuv\ and \Nivuv\ identified previously in GN-z11  \citep{Bunker2023JADES-NIRSpec-S} and subsequently in few other high-$z$ galaxies, they found 41 objects with robust detections in one or both of these UV lines. This is their definition of N-emitters, which we also adopt here, similarly to other studies. Other definitions are discussed later (Sect.~\ref{s_discuss}). 

Based on different emission line ratios and adopting simple assumptions (e.g.~strong line methods for O/H and assumed electron temperatures),  \cite{Morel2025Discovery-of-ne} have derived relative abundance ratios of C, N, and O, both with respect to H, and also directly N/O, N/C, and related ratios.  They find N-emitters spanning a wide range of metallicities (O/H) between $\sim 3$\% solar to near-solar, and consistently high N/O and N/C ratios. Typically $\log {\rm (N/O)} \sim -0.8$ to $\sim 1$, i.e.~N/O is clearly super-solar\footnote{The solar ratio is $\log($N/O)$=-0.86$, according to \cite{Asplund2009The-Chemical-Co}.}, in agreement with most studies of previously known N-emitters \citep[see][and references therein]{Ji2026Connecting-JWST,Marques-Chaves2024Extreme-N-emitt,Berg2025A-Fleeting-GLIM,Martinez2025Under-Pressure:,Moreschini2026One-cloud-is-no}.

\begin{figure}[tb]
\includegraphics[width=0.5\textwidth]{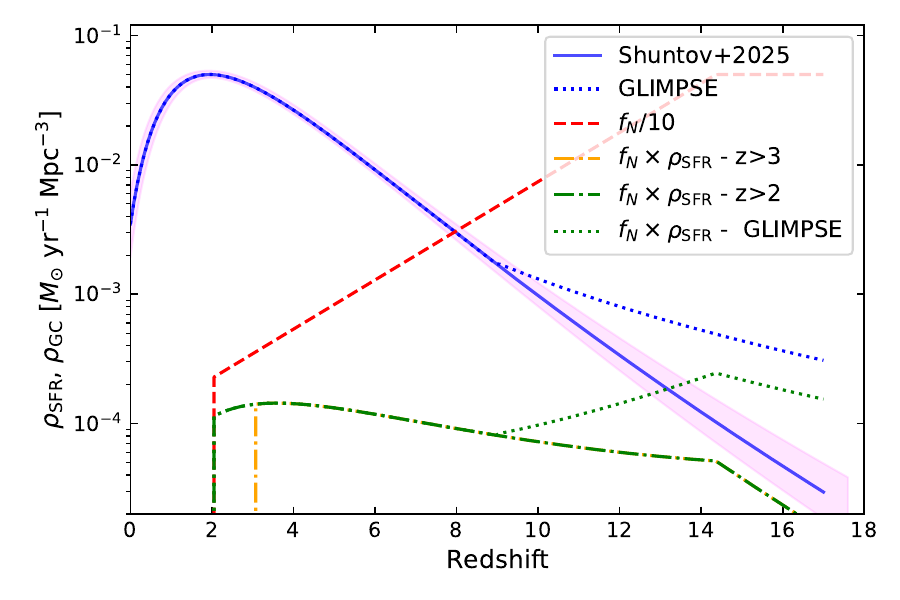}
\caption{Observed cosmic star-formation rate density evolution from \cite[][blue solid line]{Shuntov2025COSMOS-Web:-Ste} and from the GLIMPSE survey at $z>9$ \citep[][blue dotted line, after conversion to the Chabrier IMF]{Chemerynska2025The-first-GLIMP}, and predicted GC formation rate, $\rho_{\rm GC}$, as a function redshift (dash-dotted and green dotted lines). The dashed line shows the observed increase of the N-emitter fraction, $f_N$, with redshift from \cite{Morel2025Discovery-of-ne}, extrapolated down to $z=2$ and scaled by $1/10$ for convenience.}
\label{fig_sfrd}
\end{figure}

From their newly discovered N-emitters and the previously identified objects, \cite{Morel2025Discovery-of-ne} have found that the fraction of N-emitters, $f_N$, increases strongly with redshift,  when compared to all galaxies, galaxies with emission lines or objects showing just UV emission lines (``UV emitters'' in short). The observed redshift dependence trend is well described by  (see Figs.~\ref{fig_sfrd} and \ref{fig_vandels}):
\begin{equation}
\log(f_N)= 0.19 \times z -3.03,
\label{eq_fn}
\end{equation}
between $z \sim 3$ (the lower limit is set by the wavelength coverage of the JWST spectra) and $z \sim 14$, and currently undefined at higher redshift. We adopt a maximum $f_N=0.5$ (predicted at $z \ga 14.4$), and assume that the fraction of N-emitters becomes negligible at $z< \zmin \sim 2-3$. To back up this assumption, we have examined other datasets at $z \sim 2-5$ and $z \sim 0-0.4$ to provide constraints on the N-emitter fraction at lower redshift, as described in the Appendix.

If $f_N$ tracks the fraction of N-emitters among star-forming galaxies as a function of redshift, the cosmic star-formation rate density, $\rho_{\rm SFR}$, allows us to predict -- with the above assumption -- the GC formation rate, $\rho_{\rm GC}$, more precisely the stellar mass being formed in GCs per unit time and cosmic volume, as
\begin{equation}
\looseness=-1
\rhogc  = f_N \times \epsilon_{\rm GC} \times \rhosfr,
\end{equation}
where we assume $\epsilon_{\rm GC} \sim 1$ (see Sect.~\ref{s_caveat}).
The result is shown in Fig.~\ref{fig_sfrd}, where we have adopted the recent determination of $\rho_{\rm SFR}$ from \cite{Shuntov2025COSMOS-Web:-Ste}, which includes recent measurements from JWST at high-redshift, and extends down to $z \sim 0$.
Since the N-emitter fraction and \rhosfr\ both show a strong but opposite dependence on redshift, the product of the two depends only relatively weakly on redshift, predicting an increase of $\rho_{\rm GC}$ by a factor $\sim 3$ from $z \sim 16$ to $z \sim 3-4$, whereas \rhosfr\ increases by nearly 3 orders of magnitude over the same time.

To examine the impact of uncertainties on \rhosfr\ at high-$z$, we also adopt the recent result from the GLIMPSE survey, which exploits strong gravitational lensing to determine the UV luminosity function to larger depths for galaxies above $z \ga 9$ \citep{Chemerynska2025The-first-GLIMP}. While these observations agree with the compilation of \cite{Shuntov2025COSMOS-Web:-Ste} at $z \sim 9$, they indicate a slower decrease of the  star-formation rate  density towards higher redshifts. Since at $z \ga 10$ the N-emitter fraction increases rapidly above $f_N>0.1$, the predicted value of \rhogc\ could increase at these high redshifts, reaching values of $\rhogc(z=16)$ a factor $\sim 5$ higher than for the more rapidly declining \rhosfr\ history.
Although \rhogc\ depends on the product of $f_N \times \rhosfr$, in which both quantities have higher uncertainties at high redshift (see Figs.~\ref{fig_sfrd} and \ref{fig_vandels}), this has a negligible impact on the predicted age distribution of GCs and the total mass formed in GCs, which we now discuss.

\begin{figure}[tb]
\includegraphics[width=0.5\textwidth]{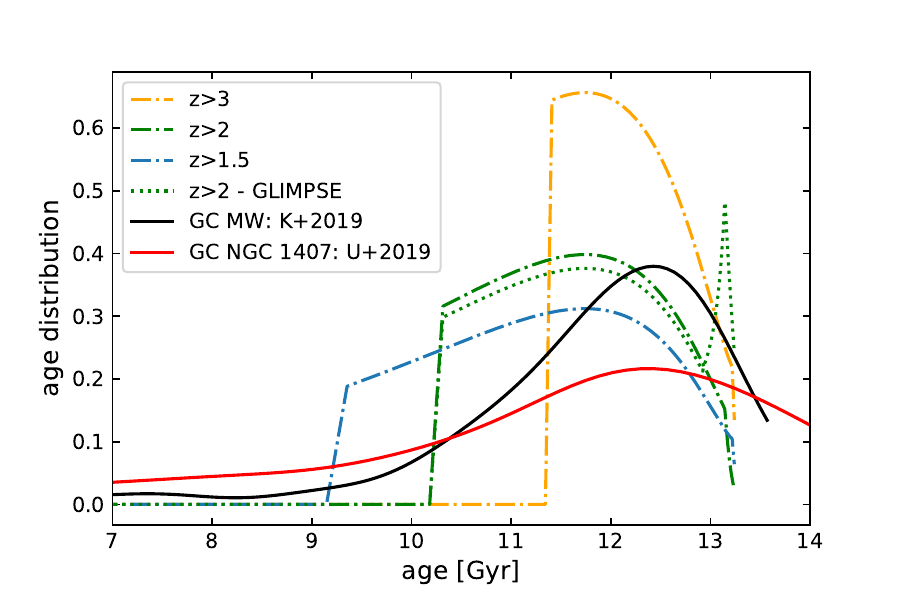}
\caption{Predicted and observed age distribution of GCs. Observational data for GCs from the Milky Way from  \cite{Kruijssen2019The-formation-a}
 and for the massive elliptical galaxy NGC 1407, from \cite{Usher2019The-SLUGGS-surv} and compiled by \cite{Valenzuela2024Galaxy-archaeol}, are shown by solid lines.  Dash-dotted lines show the predicted age distributions from our simple model, adopting the cosmic SFR density history from \cite{Shuntov2025COSMOS-Web:-Ste} for different values of $\zmin=1.5$, 2, and 3.  The dotted line shows the same $\zmin=2$ and the higher $\rhosfr(z)$ values inferred from the GLIMPSE survey. }
\label{fig_age}
\end{figure}

\subsection{Predicted age distribution of GCs}
The GC formation rate $\rho_{\rm GC}(z)$ can directly be translated to the predicted age distribution of GCs, if we assume that the mass function of GCs at their formation is constant, on average, with cosmic time. The resulting age distribution, predicted for different lower redshift limits \zmin\ of the occurrence of N-emitters is shown in Fig.~\ref{fig_age}.
In all cases the predicted distributions are asymmetric, showing a peak age for GC formation of $\sim 12-12.5$ Gyr, at rapid decrease towards older ages, and a more extended tail down to an age given by the adopted  redshift ``cutoff'' age (given by $\zmin$).  

In our simple picture, the peak age is determined by the combination of a decreasing frequency of N-emitters and the rise of cosmic SFR density,
the rapid decrease at ages $\ga 12.5$  Gyr (above the peak) is due to the decrease of \rhosfr\ at $z \ga 8$, and the tail at low ages depends on the 
behaviour of $f_N$ at $z <3$, which is currently not constrained.
In case of a slower decrease of the cosmic star-formation rate density, as suggested by the GLIMPSE survey, our simple model predicts an additional, sharp peak in the age distribution of GCs at very early ages ($\sim 13.2$ Gyr, see Fig.~\ref{fig_age}). Given the typical observational age uncertainties \footnote{For example, in the MW GC data used here from the compilation of \cite{Valenzuela2025Globular-cluste} the mean age uncertainty is 0.55 Gyr.}, this narrow peak at high-$z$ would probably remain undetected.
This also shows that the exact value of the N-emitter fraction at $z \ga 7$ is not important for the predicted age distribution and that a possible contribution of AGN to the N-emitter population at high-$z$ does not significantly affect this result. We discuss this issue further below (Sect.~\ref{s_caveat} ).
 
Observed age distributions of GC populations in the Milky Way and a massive elliptical galaxy NGC 1407 are also shown in Fig.~\ref{fig_age}, for comparison. They illustrate variations of inferred age distributions in different  galaxy types \citep[see e.g.~discussions in ][]{Valenzuela2024Galaxy-archaeol}.
Qualitatively, our predicted age distributions resemble the observed ones in all three features described above, with a peak at 12-12.5 Gyr, a tail at younger ages, and a drop towards the cosmic age. 
Compared to other existing GC formation and evolution models, including semi-analytical, cosmological hydro, and other models, our  simple models fares quite well, as can be seen from the comparison presented by \cite{Valenzuela2025Globular-cluste}. 

\begin{figure}[tb]
\includegraphics[width=0.5\textwidth]{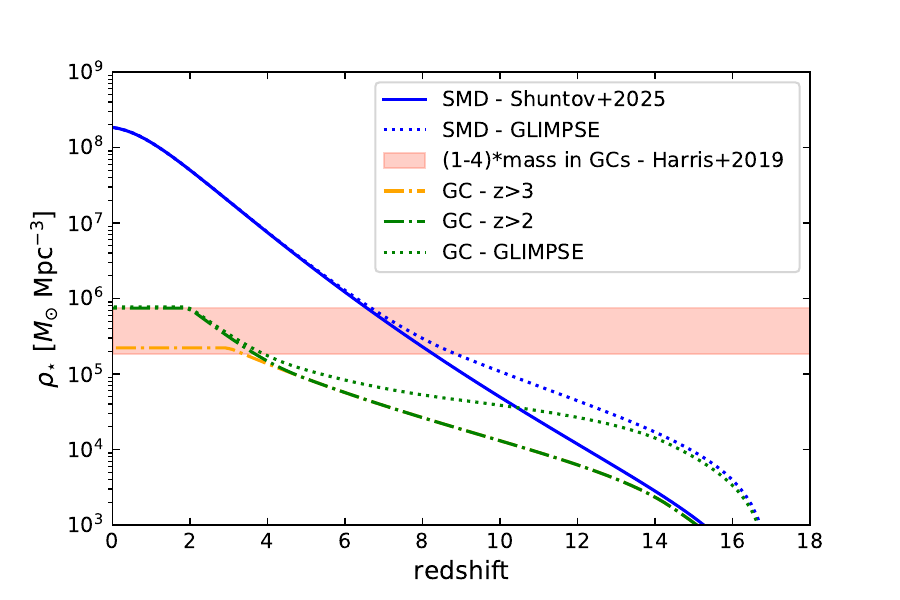}
\caption{Predicted cumulative stellar mass density evolution (SMD) in stars, \rhostar, and globular clusters, \rhostar(GC), as a function of redshift.
The SMDs are derived from the star-formation and globular cluster formation rate densities, \rhosfr\ and \rhogc, respectively, shown in Fig.~\protect\ref{fig_sfrd}. The shaded area indicates the range between $(1-4)$ times the observed cosmic stellar mass density in GCs derived at $z \sim 0$ by \cite{Harris2013A-Catalog-of-Gl}, to account for the higher initial masses of GCs.}
\label{fig_smd}
\end{figure}

\subsection{Predicted cosmic GC mass density}

Another interesting aspect of our model is that it can predict the total mass in GCs formed across cosmic time, a key quantity which can be compared to the cencus of GC populations in galaxies. To do so, we integrate the GC star formation rate density, \rhogc, over time and plot in Fig.~\ref{fig_smd} the resulting cumulative stellar mass density hosted in GC as a function of redshift. 
This is compared to the classical (cumulative) stellar mass density, \rhostar, derived from integration of the adopted \rhosfr\ from \cite{Shuntov2025COSMOS-Web:-Ste} with an adopted  constant return fraction $R=0.41$ \citep{Madau2014Cosmic-Star-For}.  At $z \approx 0$ the total stellar mass density reaches $\rhostar(z=0) \approx 2 \times 10^8$ \msun\ Mpc$^{-3}$, and our predicted mass density in globulars, $\rhostar($GC,$z=0)=(2-4) \times 10^5$ \msun\ Mpc$^{-3}$, for $2 \le \zmin \le 3$.

Observationally, numerous studies have shown strong correlations between the mass of a galaxy's  GC population and the mass of its host dark matter halo or the dynamical mass of the galaxy, across many orders of magnitude in halo or galaxy mass and for different galaxy types and environments \citep[see e.g. the review of ][]{Kruijssen2025The-Formation-o}.
\cite{Harris2013A-Catalog-of-Gl} have, e.g., determined the specific mass, $S_M$, of the GC population with respect to the baryonic mass of its host galaxy, finding a well-defined ratio of $S_M \approx 0.1$ percent.  This implies that, observationally the cosmological average at $z \sim 0$ of the GC mass density, $\rhostar($GC$) \approx 10^{-3} \times \rhostar(z=0) \approx 2 \times 10^5$ \msun\ Mpc$^{-3}$, which matches well our predictions, as shown in Fig.~\ref{fig_smd}.

Since GCs were more massive at birth, the observed present-day mass density in GCs, $\rhostar($GC,$z=0)$,  has to be corrected upward before comparing with our prediction. Various approaches indicate that typical GCs were a factor $\sim 2-4$ more massive at birth  than they are today \citep[e.g.][]{Reina-Campos2018Dynamical-clust,Baumgardt2019Mean-proper-mot}. The shaded area in Fig.~\ref{fig_smd} illustrates the range between $(1-4) \times \rhostar($GC$,z=0)$, showing that the predicted stellar mass density of GCs (at formation) is in excellent agreement with the observations in the present-day Universe.

Figure~\ref{fig_smd}  also shows that the differences between the two scenarios of the cosmic SFR evolution at high-$z$ ($z >8$) discussed above have, as expected, no significant impact on the cumulative stellar mass densities at $z \la 4$ and down to present day (z=0), since it concerns only a short fraction of cosmic time. For the same reason, uncertainties in $f_N$ at $z\ga 8$ --  due to small number statistics, ``contamination'' by AGN, or similar -- have a negligible effect on the present-day stellar mass density in GCs.

\section{Discussion}
\label{s_discuss}

As we have just shown, the hypothesis that N-emitters trace galaxies where the UV emission is dominated by young globular clusters in formation and that their  fraction follows the observed redshift  dependence identified by \cite{Morel2025Discovery-of-ne}, leads to straightforward predictions of the average age distribution of GCs (Fig.~\ref{fig_age}) and the cosmic mass density of globulars at present times (redshift $z=0$; see Fig.~\ref{fig_smd}).
Strikingly, these predictions are in very good agreement with the observations, despite the simplicity of our assumptions. 
To the best of our knowledge, this is the first empirical model  reproducing the average and distribution of the formation redshifts (or ages, equivalently) of globulars, and the total baryonic mass locked up in GCs at the present-day. 
The success of this simple model supports the idea that N-emitters could trace globular clusters, as suggested earlier, and as we now briefly discuss.

\subsection{Overall picture linking N-emitters and GCs}
The  connection between N-emitters and GCs has first been suggested based  on the resemblance of the ISM abundances of N-emitters (supersolar N/O ratios and normal C/O at low-metallicities) with GC stars  \citep{Charbonnel2023N-enhancement-i,Senchyna2024GN-z11-in-Conte,Marques-Chaves2024Extreme-N-emitt}, which are unique in showing such abundance patters and  star-to-star abundance variations  \citep[e.g.][]{Bastian2018Multiple-Stella}.
Furthermore, the compactness of N-emitters, high ISM densities ($n_e \sim 10^{4-5}$ \cmc), high mass, and high SFR surface densities indicate unusual and extreme conditions compared to other galaxies at the same redshift \citep[see e.g.][]{Marques-Chaves2024Extreme-N-emitt,Schaerer2024Discovery-of-a-,Topping2024Metal-poor-star,Arellano-Cordova2025CLASSY-XII:-nit}. This is consistent with numerous evidence showing that  GCs are only formed in rare, high-pressure environments with a high SFR surface density, and primarily at high-redshifts  \citep{Krause2020The-Physics-of-,Kruijssen2025The-Formation-o}.
Also, the abundance variations and multiple populations found in globulars, exist only above some minimum mass ($M \ga 10^4$ \msun) and compactness \citep[e.g.][]{Krause2016Gas-expulsion-i}, showing the existence of a threshold for their formation \citep{Bragaglia2017NGC-6535:-the-l,Gieles2018Concurrent-form}.
In short, the available observations suggest that N-emitters and the environments where GCs are expected to form have  all of these features/properties in common.

The  finding of  ``extreme'' conditions in N-emitters explain why these objects are relatively rare (less than 1 \% at $z \la 5.4$).
Furthermore the increasing compactness of galaxies, their increased SFR surface density and ISM density with increasing redshift,  also naturally explain the observed increase of the N-emitter fraction with $z$, as already pointed out by \cite{Morel2025Discovery-of-ne}. Finally, the existence of some threshold suggests that GC formation naturally stops or becomes more rare, in a Universe where the ISM conditions evolve. This is compatible with the extreme paucity of N-emitters at low-redshift 
\citep[see e.g.~discussions in ][]{Ji2026Connecting-JWST,Martinez2025Under-Pressure:}, which is described for simplicity by our adopted minimum redshift for N-emitters, \zmin (see Sect.~\ref{s_stats}). In Appendix \ref{s_lowz} we estimate the upper limit of the N-emitter fraction at low-redshift from a large sample of SDSS spectra.

Our simple calculations have  demonstrated quantitatively that the hypothesis that  N-emitters trace the formation of GCs is compatible with the observed age distribution and the total amount of stars formed in the GC population.  This provides further support to the association of N-emitters as signposts of GC-formation.

\subsection{Caveats and open questions}
\label{s_caveat}

We now discuss several caveats, uncertainties, and open questions related to the predictions presented here and their interpretation.

\subsubsection{Uncertainties on the N-emitter fraction and implications} 
The ``full'' uncertainty on the N-emitter fraction is not possible to establish, for multiple reasons, including since the classification of a galaxy as N-emitter (defined by the presence of one of the \Nivuv\ and \Niiiuv\ lines) depends on the detectability of these lines, hence on the quality of the spectra (depth, spectral resolution, width of the lines etc.), which is not uniform in the parent sample drawn from 8323 unique sources in the DAWN JWST Archive (DJA)\footnote{\url{https://dawn-cph.github.io/dja}}  \citep{msaexp,Heintz2025The-JWST-PRIMAL}.
 Interestingly, our determination of $f_N$ from the independent VANDELS dataset of VLT spectra of star-forming galaxies, yields consistent values of $f_N$ between $z \sim 3-5$, and confirms also the redshift increase of $f_N$ (see Appendix A), 
 supporting thus the empirical dependence of $f_N(z)$ from \cite{Morel2025Discovery-of-ne}.
 
 Should $f_N$ be systematically higher at all redshifts, e.g.~by a factor 2 as the current data may allow, this would obviously not alter the predicted GC age distribution, and simply increase \rhogc\ and the cumulative stellar mass density in GCs at all redshifts, $\rhostar({\rm GC})$, by the same amount. Such a increase can easily be compensated by lowering the GC formation efficiency, $\epsilon_{\rm GC}$, which we discuss below.
 Lower fractions of $f_N$ are very unlikely, since the N-emitter sample should be quite robust and since the total number of star-forming galaxies plus AGN is quite easily established from the parent sample, with less uncertainty than the identification of nitrogen lines.

Given different object selections and methods to determine N/O abundance ratios discussed in the literature, one may wonder how well $f_N$ used here accounts for all objects with a high N/O abundance, or if our selection misses some objects. As mentioned earlier, our UV selection (presence of \Nivuv\ or \Niiiuv) roughly translates to a supersolar N/O ratio ($\log({\rm N/O}) \ga -0.8$). Several galaxies with supersolar N/O ratios reported in the literature \citep[e.g.][]{Sanders2023A-Preview-of-JW,Stiavelli2025What-Can-We-Lea} are indeed not included in the sample of  \cite{Morel2025Discovery-of-ne}. For these objects the abundances have been derived from optical medium-resolution spectra including \Nii, but rest-UV spectra (including PRISM spectra)  are currently not available. 
However, since the number of medium-resolution spectra required to determine the N-abundance from optical spectra (\Nii) is currently small, statistical studies are not yet possible with these data, in contrast to  the large number of PRISM spectra available. Finally, for a small number of objects it has been possible to determine N/O with several methods, using separately UV and optical lines. 
In general a good or reasonable agreement between the UV and optical methods has been found \citep[e.g.][]{Berg2025A-Fleeting-GLIM,Moreschini2026One-cloud-is-no,Welch2025The-Sunburst-Ar,Marques-Chaves2024Extreme-N-emitt}, indicating that the UV-selection of N-emitters adopted by \cite{Morel2025Discovery-of-ne} should (on average) not miss objects with similarly high N/O ratios determined from optical spectra. 

Finally, the question arises if the simple measure of the N-emitter fraction misses some GCs in formation, e.g.~systems which would not or ``insignificantly'' be enhanced in nitrogen, if they exist.  For several decades now, GCs have been shown to systematically exhibit peculiar features (compared to other star clusters), which include the presence of multiple stellar populations and star-to-star variations of specific elements (N, O, Na, Al, and others) and are now the defining characteristics of globulars \citep[see e.g.~reviews of][]{Gratton2004Abundance-Varia,Bastian2018Multiple-Stella}. By definition, every GC therefore contains N-enriched stars (compared to their initial N-abundance), although with varying degrees of N-enrichment and varying fractions of stars being enriched. For GCs with masses between $10^4$ and $10^6$ \msun, e.g., the fraction of enriched stars varies from $\sim 0.4 - 0.8$ \citep{Bastian2018Multiple-Stella} and [N/Fe] increases by up to $\sim 1.3$ dex \citep[e.g.][]{Carlos2023The-chemical-co}. To summarise, at the time of formation of the enriched stars, all GCs must have had some N-enriched gas. How this relates with the enrichment seen in the ionised ISM can, however, not be quantified.  In practice this means that our method could miss some GCs, a case  already discussed above.

\subsubsection{Contamination by AGN ?} 
Despite the various and new arguments in favour of a link between N-emitters and young globular clusters, the nature of N-emitters remains debated.
\cite{Zhu2025The-Nature-of-N} have recently studied eight  N-emitters at $z \sim 6-11$ with medium-resolution spectra from JWST and conclude that seven of them are best described by N-enhanced AGN models. The question hence arises whether our conclusions would be affected, if the majority of N-emitters at $z \ga 6-8$ would be AGN.
In fact, the AGN classification remains also debated, as illustrated by  two well-studied N-emitters where no consensus has been reached, GNz-11 and CEERS-1019 \citep{Larson2023A-CEERS-Discove,Maiolino2024A-small-and-vig,Alvarez-Marquez2025Insight-into-th,Zamora2025GA-NIFS:-Unders}. In addition, most, if not all N-emitters studied so far show compact but resolved morphologies \citep[cf.][]{Morel2025Discovery-of-ne}. This suggests that AGN are unlikely to be the dominant source of the UV continuum in N-emitters.
Furthermore, if the fraction $f_N$ of N-emitters associated with star- or GC-formation would be reduced at high-$z$, this has a  small or negligible impact on the predicted GC age distribution and the total mass formed in N-emitters, as shown above.
Finally,  analysing the largest N-emitter sample known to date, \cite{Morel2025Discovery-of-ne} found only a small fraction of objects ($4/45 \approx 8$\%) with signatures of broad-line AGN, and few such objects at $z \la 7$.  Further work is needed to establish the importance of AGN among N-emitters and test the various proposed scenarios.
 In any case, our main results remain unchanged if the majority of N-emitters at $z \la 7$ are related to star cluster formation, not AGN.
 
 \subsubsection{High GC formation efficiency } 
In our calculation of the total stellar mass density in GCs, \rhostar(GC), we assume that the bulk of star-formation in N-emitters produces (and ends up in) globular clusters, i.e.~$\epsilon_{\rm GC} \sim 1$, which is probably a somewhat optimistic assumption. This basically postulates a high cluster formation efficiency (CFE), and a cluster mass distribution with a minimum cluster mass  close to or above the threshold for the formation of GC ($\sim 10^4$ \msun), or in other words, a small amount of mass in low-mass clusters which cannot become globulars.

Whereas high CFE$\sim 0.5-1$ are commonly observed in the high-$z$ Universe and in high SFR surface density galaxies \citep[e.g.][]{Adamo2020Star-cluster-fo}, the mass function of stellar clumps or clusters is currently not yet constrained at masses below $\sim 10^5$ \msun\  \citep{Claeyssens2025Tracing-star-fo}. 
In fact, one case of a high-$z$ galaxy showing a CFE near unity and a cluster mass function leaving no room for low-mass clusters, has recently been found by  \cite{Vanzella2025The-z--9.625-Co}; it is nick-named the ``Cosmic Gems'' galaxy. Whether this is a common property remains to be explored in the future.
 Although maximal, these assumptions are therefore not incompatible with current observations, and it should be reminded that they only need to apply to rare objects (N-emitters), which already show other properties which are ``extreme'' (cf.~above).
 
We  also note that the N-emitter fraction adopted might be underestimated, since it relies on the detection of UV emission lines, which is biased towards strong N lines \citep{Zhu2025Only-Nitrogen-E}. This, or other cases leading to an underestimate of $f_N$, would allow us to relax the assumption of $\epsilon_{\rm GC} \sim 1$ and maintain a total stellar mass density in GCs compatible with observations at $z \sim 0$. Physically, values of $\epsilon_{\rm GC} <1$ are easy to explain, e.g.~due to the formation of clusters below the mass limit of GCs or due to the evaporation or tidal destruction of globulars \citep[see e.g.~references in ][]{Kruijssen2025The-Formation-o}.

Future investigations will be necessary to firm up or infirm the suggested  link between N-emitters and proto-GCs, to which our approach adds a new statistical aspect on their populations. 

\section{Conclusions}
\label{s_conclude}
It has been proposed that the so-called N-emitters, strongly N-enhanced galaxies with uncommon UV emission lines of nitrogen revealed by JWST observations, could be related to young globular clusters (GCs) forming in situ \citep{Charbonnel2023N-enhancement-i,Senchyna2024GN-z11-in-Conte}. 
To test this hypothesis we have used the recently determined frequency of N-emitters and their observed redshift evolution from  \cite{Morel2025Discovery-of-ne} to compute the cosmological average of the GC star-formation rate density, \rhogc, the resulting age distribution of globulars, and -- to the best of our knowledge -- for the first time the cumulative stellar mass density in GCs across cosmic times, from the observed cosmic star-formation history $\rhosfr(z)$.

We found that GC formation is predicted to peak at $z \sim 3-4$  (ages $\sim 11.5-12$ Gyr), with an asymmetric age distribution showing an extended tail to younger ages and a rapid drop beyond $\ga 12.5$ Gyr, quite independently of the exact values of \rhosfr\ at very high redshifts, where it is currently being determined.  The predicted GC age distribution resembles the typical distribution of GCs in the Milky Way and in elliptical galaxies.

The predicted total stellar mass density formed in N-emitters down to redshift zero, $\rhostar($GC,$z=0)=(2-4) \times 10^5$ \msun\ Mpc$^{-3}$, matches within a factor $\sim 2$ the observed fraction of stellar mass found in the GC population at $z=0$ \citep[$=10^{-3}$ of the total stellar mass density of galaxies,][]{Harris2013A-Catalog-of-Gl}.
Our results provide additional support to the hypothesis that N-emitters represent signposts of GC formation, and that these phenomena are thus related.

\begin{acknowledgements}
DS wishes to thank the IAP, Paris, and its staff for their hospitality during a stay where some of this work was done.
We also thank Lucas Valenzuela for sharing results in electronic format.
YI and NG acknowledge support from the National Academy of Sciences of Ukraine (Project No. 0126U000353).
\end{acknowledgements}

\bibliographystyle{aa}
\bibliography{}

\begin{appendix}
\label{s_app}
\section{New constraints on the N-emitter fraction at $z \sim 2-5$ from VANDELS}

To complement and possibly extend the recent statistics of N-emitters derived from JWST spectra to lower redshifts, we have examined the data from VANDELS, a deep, uniform, spectroscopic galaxy survey undertaken with the VIMOS instrument at the VLT. To do so, we have used the spectra from the latest data release \citep{Garilli2021The-VANDELS-ESO} and, following  \cite{Talia2023The-VANDELS-ESO}, we have extended the catalog of spectroscopic measurements to include the  \Niiiuv\ and \Nivuvab\  features. This is done with Gaussian fits and direct integrations adapted to the doublet or multiplet structure of these line complexes\footnote{This follows the methods from \cite{Schreiber2018Near-infrared-s} and \cite{Borghi2022Toward-a-Better}.}.

Eliminating passive galaxies from the full sample, following the UVJ selection described in  \cite{Garilli2021The-VANDELS-ESO},  we are left with 1501 galaxies in the redshift interval $1.75 \la z \la 5.75$, where \Niiiuv\ or \Nivuvab\ can in principle be detected. 
Our automated measurement yields $\sim 10$ candidates at $\ga 3-4 \sigma$, including the well-known N-emitter VANDELS\_CDFS\_003073, an AGN at $z=5.56$ \citep[see][and references therein]{Barchiesi2023The-ALPINE-ALMA,Ji2026Connecting-JWST}.
After visual inspection, only one additional object with a credible \Niiiuv\ detection, VANDELS\_CDFS\_247555 at $z=3.745$ remains, yielding thus a total detection fraction 
 of N-emitters of $1.3 \times 10^{-3}$  among the VANDELS star-forming galaxies and AGN showing rest-UV emission.
 
\begin{figure}[htb]
\includegraphics[width=0.5\textwidth]{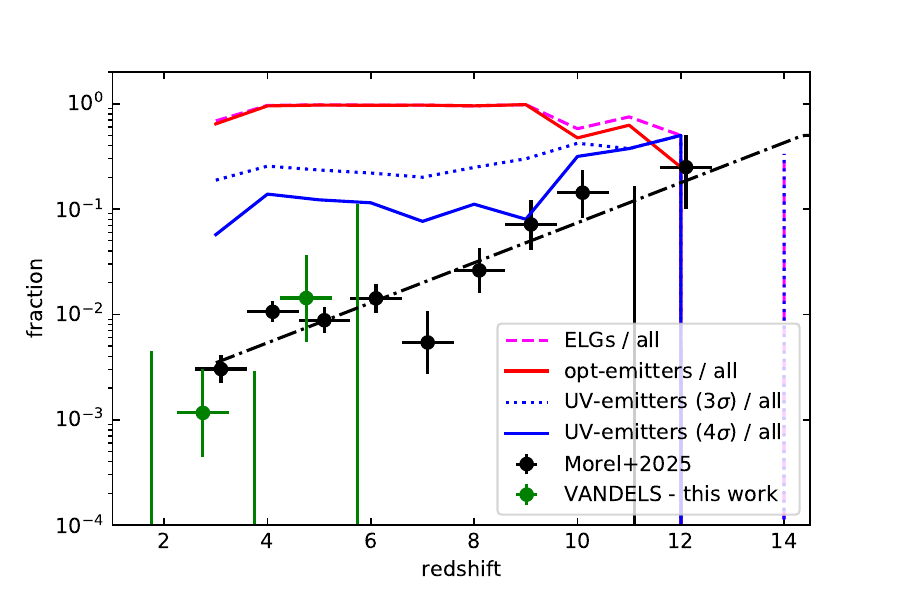}
\caption{Redshift evolution of the fraction of N-emitters (circles) as a function of redshift, determined by \cite{Morel2025Discovery-of-ne} from the JWST NIRSpec PRISM spectra (black circles and limits) and from the VANDELS survey (green circles and limits). Uncertainties show the 68\% confidence range of the fractions, and the bin width along the x-axis.
Blue lines show the fraction of rest-UV line emitters $4 \sigma$ (solid) or $3 \sigma$ detection (dotted) thresholds, respectively. The fraction of objects showing rest-optical emission lines is shown by the red line, and all emission-line galaxies (UV and/or optical) in magenta (dashed). 
The black dash-dotted line shows the fit to the $z>3$ data, Eq.~\ref{eq_fn}, obtained by \cite{Morel2025Discovery-of-ne}.
}
\label{fig_vandels}
\end{figure}

 From these detections and non-detections we derive the fraction of N-emitters in several redshift bins between $1.75 \la z \la 5.75$, computing the confidence intervals with the Wilson interval formula, which is adapted for small sample sizes and low fractions.
 The result is plotted in Fig.~\ref{fig_vandels} together with the N-emitter fraction determined by \cite{Morel2025Discovery-of-ne} from archival JWST NIRSpec PRISM data including large samples.
 Clearly, the VANDELS and JWST data are consistent with the reported redshift evolution of $f_N$ (Eq.~\ref{eq_fn}). Furthermore, the VANDELS data is compatible with a low fraction at $z \sim 2-3$, although the statistical uncertainties are relatively large.  The new data thus supports the adopted description of $f_N(z)$ and  drop at $\zmin \sim 2-3$.
 
 Finally, we note that the $f_N$ values obtained from the JWST and the VANDELS data agree approximately, despite the different properties (spectral resolution, average S/N etc.) of the data. Indeed, while the typical threshold (3 $\sigma$ upper limits) for the \Niiiuv\ and \Nivuv\ lines is $\sim 10$ \AA\ in the JWST PRISM spectra \citep[see][]{Morel2025Discovery-of-ne}, the median 3-$\sigma$ limit  is $\sim 6$ \AA\ in the VANDELS spectra, i.e.~somewhat but not significantly lower.
 
\section{On the N-emitter fraction at low redshift}
\label{s_lowz}

\begin{figure}[htb]
\includegraphics[width=0.5\textwidth]{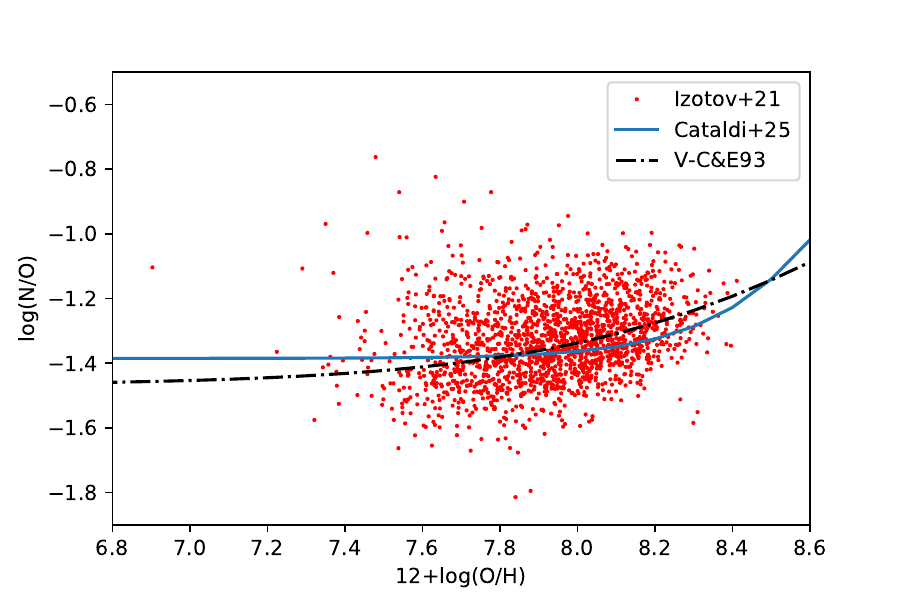}
\caption{ N/O abundance ratio as a function of O/H from the sample of 2212 low-$z$ compact star-forming galaxies from \cite{Izotov2021Low-redshift-co} with  \Oiiit\ detections. N/O was determined from the  \Nii/\Oii\ ratio using the recent calibration from \cite{Cataldi2025Tracing-Nitroge} for low-$z$ galaxies. The solid line shows their mean relation at low-$z$, the dash-dotted line the one from  \cite{Vila-Costas1993The-nitrogen-to}.  
The figure illustrates the absence of low-$z$ objects with very high N/O abundances ($\log({\rm N/O}) \protect\ga -0.8$), comparable to those of the objects classified as N-emitters in \cite{Morel2025Discovery-of-ne} and this work.
}
\label{fig_no_oh_csfg}
\end{figure}

To the best of our knowledge, no systematic search for the \Nivuv\ and \Niiiuv\ lines has so far been undertaken in galaxy spectra at low-$z$, and the current compilations of galaxies with UV spectroscopy from the Hubble Space telescope (HST) and earlier, are small \citep[see e.g.][who cite other compilations, including in total 270 galaxies]{Berg2022The-COS-Legacy-}. Focussing  on 46 high-quality HST spectra primarily from the CLASSY survey \citep{Berg2022The-COS-Legacy-}, \cite{Martinez2025Under-Pressure:}  have recently  identified up to 8 galaxies with clear or possible nitrogen emission lines, including the well-studied WR-galaxy Mrk 996 (J0127-06719), which was previously reported as a clear N-emitter at low-$z$ \citep{Thuan1996Hubble-Space-Te}. From their abundance determinations using UV and optical emission lines, \cite{Martinez2025Under-Pressure:}  find only one object (Mrk 996) with a robust super-solar N/O ratio. 

Mrk 996 has long been known as a blue compact dwarf galaxy (BCD) with very peculiar properties, showing a very compact nucleus, broad emission lines, WR features, very high electron densities in the nucleus, and other unusual properties \citep[e.g.][]{Izotov1994The-Primordial-,Thuan1996Hubble-Space-Te}. Furthermore, it was found among a small group of BCDs with a significant overabundance of N/O  compared to other galaxies and BCDs at similar metallicities \citep[O/H; see e.g.][]{Pustilnik2004HS-08374717---a,James2009A-VLT-VIMOS-stu}. 

Larger samples including the SDSS, have not found significant numbers of objects with high N/O abundances, comparable to those of the high-$z$ N-emitters. 
To quantify this, we have used the spectroscopic sample of $\sim 25000$ compact star-forming galaxies constructed by \cite{Izotov2021Low-redshift-co} who have used SDSS spectra. Among those, 2212 objects have a significant \Oiiit\ detection, allowing for O/H measurements using the direct method. For this subsample we then determined the N/O abundance ratio from the \Nii/\Oii\ ratio, using the recent calibration from \cite{Cataldi2025Tracing-Nitroge}.The result is illustrated  in Fig.~\ref{fig_no_oh_csfg}. We find a mean N/O ratio of  $\log({\rm N/O})= -1.36^{+0.14}_{-0.23}$, and a slight dependence of N/O with O/H, as well known from the literature. Most importantly, very few or no objects are found at the high N/O ratios found in N-emitters $\log({\rm N/O}) \ga -0.8$, which translates to an upper limit of $\la 4.5 \times 10^{-4}$ for the fraction of these objects among the sample of galaxies with accurate O/H at $z \la 0.4$. 

For comparison, \cite{Bhattacharya2025The-origin-of-e} recently determined the N/O abundances for a sample of 944 star-forming galaxies with direct abundance determinations using spectra from the Dark Energy Spectroscopic Instrument (DESI). 
Among those, they report the finding  of 19 galaxies with  $\log({\rm N/O}) \ge -1.1$, which they denote as ``extreme N-emitters'', and four objects with super-solar N/O ratios at $\oh <7.5$. 
Inspection of the spectra of these four objects shows that the  \Oiiit\ line is not securely detected and/or that their oxygen abundances are unlikely to be very low since most of these objects have relatively high stellar masses, including one large spiral galaxy (their Fig.~4), which appears at odds with the very low metallicities reported for these systems.
From this we conclude that their sample is compatible with the above analysis of the larger SDSS sample, and that essentially none of these galaxies show  N/O abundance ratios comparable to the N-emitters, as defined in this work. 
Assuming that  the N/O abundance ratios determined from optical lines (\Nii\ and \Oii) agree on average with those from the UV lines and that N-emitters have typically $\log({\rm N/O}) \ga -0.8$, we therefore estimate  an upper limit of the N-emitter fraction of $f_N(z \sim 0-0.4) \la  4.5 \times 10^{-4}$ at low-$z$, from the currently largest datasets available. This limit supports our assumption of a rapid drop of $f_N$ at $z \la 2$.

\end{appendix}

\end{document}